\documentclass[letterpaper,aps,prl,twocolumn,amsmath,showpacs,amssymb]{revtex4}
\usepackage{color}
\usepackage[pdftex]{graphicx,hyperref}
\usepackage{dcolumn}
\usepackage{bm}
\usepackage{marvosym}
\usepackage{times,amsmath,amssymb}
\begin{document}

\title{Probing spatial correlations in a system of polarizable particles via measuring its optical extinction spectrum}

\author{R. M. S. Pereira, P. Pereira}

\affiliation{Centro de Matem\'atica, Universidade do Minho,
Campus de Gualtar, Braga 4710-057, Portugal}

\author{G. Smirnov, M. I. Vasilevskiy}

\affiliation{Centro de F\'{\i}sica, Universidade do Minho,
Campus de Gualtar, Braga 4710-057, Portugal}

\begin{abstract} It is shown that quantitative information on spatial correlations in a system of polarizable particles can be extracted directly from its experimentally measurable optical spectra. For a collection of metallic nanoparticles (NPs), it is demonstrated that the degree of short-range correlation in NP's positions can be evaluated by an appropriate numerical analysis of the extinction spectrum in the surface plasmon resonance region, given the polarizability of an individual NP. The spectrum analysis consists in the evaluation of a single number, which is the derivative of the ensemble response function in the vanishing polarizability limit, to be compared to pre-calculated values for a model NP system with given density and correlation parameters. 

\end{abstract}


\maketitle

{\it Introduction.}
The optical properties of assemblies of almost identical polarizable particles, randomly distributed in space and coupled by electromagnetic interactions have attracted attention for a long time \cite{Bohren-Huffman,Shalaev}, since they are of interest for fields ranging from interstellar dust \cite{Draine} to electrorheological fluids \cite{Halsey} to biosensors and single molecule detectors \cite{Freeman,Stockman}. Metallic nanoparticles (NPs) dispersed in a dielectric matrix are a practically important example of such assemblies \cite{Blackman}. These nanocomposite materials have been proposed for environment sensing \cite{Anker,Chegel}, tailoring color of functional coatings \cite{Torell}, enhancement of the non-linear response \cite {Hache,Stockman94}, energy harvesting in solar cells \cite {Ferry} and Raman scattering by molecules \cite{Kim}, amongst other applications of the nanophotonics. Theoretical description of the optical properties of a composite containing NPs becomes particularly interesting (and complex) when the polarizable particles are non-randomly distributed in space \cite{Hui,Markel,Vasilevskiy96}, e.g. aggregate into clusters, possibly of a fractal dimension \cite{Markel,Curtin}. Note that NP aggregation is precisely one of the mechanisms proposed for sensing \cite{Stockman,Chegel}; also it can enhance the detection capability of surface-enhanced spectroscopy methods \cite{Chumanov}.

Is it possible to extract information concerning the spatial correlations in
the particles' positions from the experimentally measurable optical response of the system without performing tedious fitting of the spectra using a physically founded numerical model? In the present work, we address
this fundamental question by studying ensembles of nanospheres described by so called coupled dipole equations (CDE) \cite{Shalaev,Markel,Vasilevskiy96}. CDE are linear
algebraic equations relating the polarization of each particle to the external electric field. The
dipole-dipole interactions between the particles are represented by a second rank tensor and CDE
contain a spectral variable and some parameters depending on the material properties. The purpose of this work is to study the relation between the
spectral-dependent solutions of the CDE and the type and parameters of the particle-particle correlation
function in real space. It may seem desperate trying to solve the inverse problem in order to obtain the particle-particle correlation function from an experimentally measurable spectrum of some quantity, such as absorption or extinction, because the spatial correlations affect the collective optical properties in a rather indirect way. However, one may hope to solve a more modest but still relevant problem consisting in the use of the optical spectroscopy data in order to distinguish between some classes of spatial distributions of the particles, e.g. between a completely random distribution and a short-range correlated one where the particles form closely-spaced pairs. Our calculations are performed for a collection of metallic NPs (with the dielectric
function described by the Drude model), distributed over a square or a cubic lattice. We will show that it is possible to propose a numerical criterium that allows for conclusions concerning the degree of short-range correlation in the particles' positions using the (experimentally measurable) extinction spectrum of the NP system. The central result consists in the calibration of this criterium as a function of two arguments, NP concentration and short-range correlation parameter, thus making it possible to classify an ensemble of NPs according to their collective optical response. 
Therefore one can hope that, at least, some classes of non-random distributions of the NPs can be distinguished by means of a rather simple analysis of their extinction spectra.




{\it Coupled Dipole Equations.} First, we present the formalism used to describe the electromagnetic (EM) interactions between polarizable particles that randomly occupy some fraction of the sites of a regular lattice. If the characteristic wavelengths of the EM interactions involved are much larger than the size of each NP and the distance between them, the CDE system can be written as \cite{Shalaev}:
\begin{equation}
\label{eq:1}
\mathbf{d}_i=\alpha_0 \left(\mathbf{E}_0+\sum_{i \neq j}\mathbf {T}_{ij} \mathbf{d}_j \right),
\end{equation}
where $i$ and $j$ run over all $N$ lattice sites, $\mathbf{d}_i$ is the dipole moment located at site $i$, $\mathbf{E}_0$ is an external electric field that will be taken as uniform and directed along the $z$ axis in 3D, $\mathbf{E}_0=\left ( 0,\:0,\:E_0 \right )$, where $ E_0$ is a constant. In Eq. (\ref {eq:1}) $\mathbf {T}_{ij}$ is the dipole-dipole interaction tensor,
$T_{ij}^{\alpha \beta}=R_{ij}^{-3}(3n_{\alpha}n_{\beta}-\delta_{\alpha\beta})$, where $R_{ij}$ is the distance between lattice sites $i$ and $j$, $\alpha$ and $\beta$ denote Cartesian components and $\mathbf{n}$ is a unitary vector in the direction from site $i$ to site $j$.  Let us express it in units of the lattice constant, $l$, so $T_{ij}$ is dimensionless and
 corresponds to a quasistatic approximation, in which $T_{ij}$ is a real and symmetric matrix. The NP's polarizability, $\alpha_0$, is a complex function of frequency, $\omega$. It is an experimentally controllable parameter. If the dipoles are spherical metallic (nano) particles, then we have
$\alpha_0 (\omega)=a^3 ({\epsilon(\omega)-1})/({\epsilon(\omega)+2})$, and, within the classical Drude model \cite{Blackman}, $\epsilon(\omega)=(\epsilon_{\infty}/\epsilon_{h})\left (1-{\omega_p^2}/({\omega(\omega +i \Gamma_p))}\right )$.  Here, $a$ is the NP radius, $\epsilon_{h}$ is the dielectric constant of the host matrix, and $\epsilon_{\infty}$, $\omega_p$ and $\Gamma_p$ are the parameters of the plasma resonance in the NP material.  

Let $\cal D$ be the dimension of the space (${\cal D}=3$ if we are dealing with a 3D problem  and ${\cal D}=2$ in the 2D case). 
Denote by  $|p\rangle$  a vector with ${\cal D}cN$  components ($c$ is the fraction of lattice sites occuppied by the particles). Let $I$ be the set of indices corresponding to the occupied sites. Then we have
 $|p\rangle=(p_{i_1x},p_{i_1y},p_{i_1z},p_{i_2x},p_{i_2y},p_{i_2z},..., p_{i_{{\cal D}cN}x},p_{i_{{\cal D}cN}y},p_{i_{{\cal D}cN}z})$, where the index $i_k$ is included if and only  if it belongs to $I$. We define the dimensionless polarization vector components, $p_{i\alpha}$, as
$p_{i\alpha}={d_{i\alpha}/(}{E_0 \alpha_0})$, $i\in I$.  The constant vector $|e \rangle $ has components given by
$e_{i\alpha}= e_{\alpha}=(0,1)$,  $i\in I$, if ${\cal D}=2$, and,
$e_{i\alpha}= e_{\alpha}=(0,0,1)$, $i\in I$, if ${\cal D}=3$.
By $\widehat{M}$ we denote the ${\cal D}cN\times {\cal D}cN$-matrix obtained from the ${\cal D}N\times {\cal D}N$ matrix formed by blocks $T_{ij}$, $i,j=\overline{1,N}$, 
eliminating all zero rows and columns. This is a dimensionless symmetric matrix with zero diagonal elements.

\begin{figure}
\includegraphics[width=8cm]{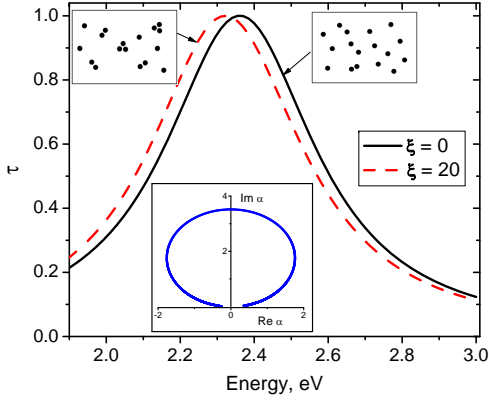}
\caption{(Colour online) Ensemble response function spectra, normalized to the maximum value, $\tau =\phi (\alpha _0 (\omega ))/\max [\phi ]$), for random ($\xi =0$, full curve) and strongly clustered ($\xi =20$, dashed curve) distributions of Au NPs on square lattice ($c=0.2, \: a/l=0.25$). SPR mode of a single NP is located at 2.37 eV, the peak at 2.31 eV is due to pairs of NPs occupying sites that are nearest neighbours along the direction of the applied field. The inset shows the hodograph of the function $\alpha _0$ for a Drude particle with $\omega _p=6.91$ eV, $\epsilon _\infty=1.53$ (bulk Au parameters), $\Gamma _p$ calculated as in Ref. \cite{Torell} and $\epsilon _{h}=5.75$.} 
\label{tran:fig2}
\end{figure}

Now we can formulate the problem in the matrix form. Equation (\ref{eq:1}) is rewritten as
\begin{equation}
\label{eq:6}
|p\rangle =|e\rangle +\alpha_0 \widehat{M} |p\rangle \:.
\end{equation}
Its solution has the form $|p \rangle =(\widehat{I}-\alpha_0 \widehat{M})^{-1} |e \rangle \equiv \widehat G |e \rangle$, 
where $\widehat G$ is Green's function. If we denote the eigenvalues of $\widehat{M}$ by $g_k\:(k=1,2...,{\cal D}cN)$ and the corresponding eigenvectors by $|u_k \rangle$, then
$\widehat G \left (\alpha _0(\omega )\right )=\sum_{k=1}^{{\cal D}cN} {|u_k \rangle \langle u_k|}({1-\alpha_0 g_k})^{-1}$. 
It is a ${\cal D}cN\times {\cal D}cN$ matrix depending on the spectral variable. Although normally the EM frequency ($\omega $) is used as such, we may also consider the polarizability 
($\alpha _0$) as a complex spectral variable.

Most observable quantities of interest can be expressed via the Green function. One of these, the extinction cross section characterizing the processes of both absorption and scattering of light by the particles, is given by \cite {Born}:
\begin{equation}
\sigma_e (\omega )=\frac  {8\pi ^2} {\lambda} \Im m \left ( \alpha_0 \sum_{n=1}^{N_1} \langle e|p \rangle _n F(\vec {\delta}_n)\right)\:,
\label{extinction}
\end{equation}
where $\lambda $ denotes the wavelength, $F(\vec {\delta}_n)$ is the distribution function of $N_1$ possible configurations of the $N$-dimensional vector, $\vec \delta =(\dots,\;\!\delta _i,\!\dots)$, that satisfy the condition, $|\vec \delta |=cN$. The number $N_1$ is given by
$N_1= {N!}/({(cN)![(1-c)N]!})$, $\sum_{n=1}^{N_1}F(\vec {\delta}_n)=1$, and  $\sum_{n=1}^{N_1}\delta_i F(\vec {\delta}_n)=c$, for any site $i$.
The scalar product$\langle e|p \rangle _n$ can be expressed through the Green function of the corresponding configuration of NPs:
$\langle e|p \rangle _n=\left \langle e \left |\widehat G^{(n)}\right |e \right \rangle$. 
Thus, the calculation of the optical response of a system of coupled dipoles with known $F(\vec {\delta}_n)$ is a straightforward (although possibly computationally hard) task. Two examples of calculated response spectra are shown in Fig. \ref{tran:fig2}. NP pairs oriented parallel to the applied field produce the low energy peak in the strong clustering case, as it has been observed experimentally \cite {Rechberger2003}.



{\it Ensemble Response Function.} Our purpose is to define a quantity that is connected with the distribution function, $F(\vec {\delta}_n)$, and, at the same time, directly related to the spectral properties of $\widehat{G}$.    
Let us consider the function
\begin{eqnarray}
\phi (\alpha )=\sum _{n=1}^{N_1}{ \left\langle e \left\vert \widehat G^{(n)}(\alpha )\right \vert e \right \rangle  F(\vec{\delta}_n)}\:.
\label{phi}
\end{eqnarray}
The values of $\phi (\alpha )$, which we shall call ensemble response function (ERF), are known for $\alpha =\alpha _0(\omega )$ [i.e. on the hodograph shown in the inset of Fig. \ref {tran:fig2}], since they can be obtained from experimental data for the extinction cross-section.

Let $\omega_*$ be such that  $\mathrm{Re}(\alpha_0)(\omega_* )=0$. Set $\alpha_*= \alpha_0(\omega_* )$.  The function $\phi (\cdot )$ is analytic in a disk $\{ \alpha\mid |\alpha-\alpha_*|< \sqrt{\alpha_*^2+\Delta ^2}; \:\Delta >0 ,\:\Delta <<1\}$ containing zero. In a small neighbourhood of zero we have:
\begin{eqnarray}
\phi (\alpha )=
\sum _{n=1}^{N_1}{ \left [ \left \langle e \left \vert e \right. \right \rangle  F(\vec{\delta}_n)
+\alpha \left \langle e \left \vert \widehat M^{(n)} \right \vert e \right \rangle  F(\vec{\delta}_n)+\dots \right ] }\nonumber\\
=Nc+\alpha \sum_{i,j}K_{ij}\sum _{\alpha ,\!\beta}e_\alpha e_\beta T_{ij}^{\alpha \beta}+\ldots
\label{phi}
\end{eqnarray}
We are interested in cases where the (random) occupation of the lattice sites is correlated. This is described by the particle-particle correlation matrix $K_{ij}$ defined as follows:
$K_{ij}=\sum_{n=1}^{N_1} \delta_i \delta_j F(\vec {\delta}_n)$. Then it follows from Eq. (\ref {phi}) that the ERF derivative taken at $\alpha = 0$ is directly related to the correlation matrix, 
\begin{equation}
\phi ^\prime(0)=\sum_{i,j}K_{ij}\sum _{\alpha ,\!\beta} e_\alpha e_\beta T_{ij}^{\alpha \beta}\:.
\label {phi'}
\end{equation}

The ERF derivative value (\ref {phi'}), with an appropriate normalization, is the proposed criterium for the evaluation of the degree of particle-particle correlations in the NP ensemble. Obviously, it is not possible to extract the correlation matrix elements from the optical response spectrum, not even the pair correlation function, $k (r) = \sum _{j=1}^N \sum _{(i>j;\:R_{ij}=r)} K_{ij}$ (where $r$ is the distance between the 1-st, 2-nd, {\it etc } nearest neighbors in the lattice).
This single number, obtainable from an experimentally measured spectrum, allows one to classify the NP ensemble as e.g. a strongly aggregated or a gas-like uncorrelated system. Of course, one needs calibration of the possible values of this quantity, which can be obtained by solving CDEs for a set of model systems where the particles are distributed according to predefined rules. Then the ERF derivative value obtained from the spectrum of the {\it real} system will show its similarity to one of these models. This is much less time consuming than the fitting of the experimental spectra.       

So, technically the problem is reduced to the evaluation of $\phi ^\prime(0)$ from the known values of $\phi (\alpha _0(\omega ) )$. It can be done by means of an appropriate extrapolation procedure that will be used for both real and model spectra. For {\it real} spectra, this is a hard ill-posed problem (note that, according to Eq. (\ref{extinction}), the observable quantity that will allow us to compare simulation results with reality is proportional to $\Im m (\alpha_0 \phi(\alpha_0))$).
However, since the function $\phi$ is a rational function with poles of $\phi (\alpha )$ concentrated in the real axis, the Tikhonov regularization method \cite {Tikhonov,Morozov} gives good results.   


{\it Examples.} For a 3D lattice and $z$ axis chosen along the direction of the external field, $\phi ^\prime(0)=\sum_{i,j}K_{ij} T_{ij}^{zz}$. If the lattice is cubic and the particle distribution is isotropic, we can write in the continuum  approximation:
$$
\phi ^\prime(0)=\int {k(r)\frac {1-3\cos ^2\theta}{r^3}r^2 d\Omega dr}\:.
$$
The integration yields zero whatever the pair correlation function is, since it depends only on the modulus of the radius--vector. Therefore, for an isotropic 3D ensemble of spherical particles, the ERF derivative is equal to zero. This is an important result, although it excludes the possibility of probing the correlations in this case.  

Let us consider a square lattice, which can be seen as an ultimate anisotropic limit of the 3D case. Nearly 2D structures can be built from metal NPs grown by colloidal chemistry techniques \cite {Freeman,Chumanov,Marzan2001,Langmuir2002,Fritzsche1997}. Aggregation of NPs in such layers can occur as a result of some chemical treatment \cite {Fritzsche1997}. We shall limit ourselves by the case of short-range correlations that can be described by a single short-range order (SRO) parameter, $\xi $. If one particle is located at a lattice site $i_1$, the relative occupation probabilities by another particle for the remaining $N-1$ sites ($i\neq i_1$) are given by: 
$P_{i /i_1}=c+\Delta_{ii_1} \xi$, where $\Delta_{i\:i_1}=1$ if $ R_{ii_1}=1$, and $ \Delta_{ii_1}=0$ otherwise. If $\xi >0$, this distribution corresponds to NP clustering. Using the Monte Carlo (MC) method, we generated 
a sufficiently large number of samples containing $cN$ particles distributed over $N=N_x^2$ sites of a square lattice, and calculated the ERF spectra [by solving Eqs. (\ref {eq:1})] and the correlation matrix $\langle K_{ij} \rangle$, for given values of $\xi $ and $c$. Two examples of the calculated spectra are shown in Fig. \ref {tran:fig2}. Since the eigenmodes of the system (for each MC sample) are known in this case, the {\it normalized} ERF derivative, $\tau ^\prime(0)$, could be calculated explicitly and MC-averaged, as a function of $c$ and $\xi $. As we established empirically, the values of $\tau ^\prime(0)$ become almost independent of the lattice size for $N_x>20$. Of course, the results depend on the single particle polarizability and its definition requires the knowledge of several essential parameters to be obtained from the experiment. 

\begin{figure}
\includegraphics[width=8cm]{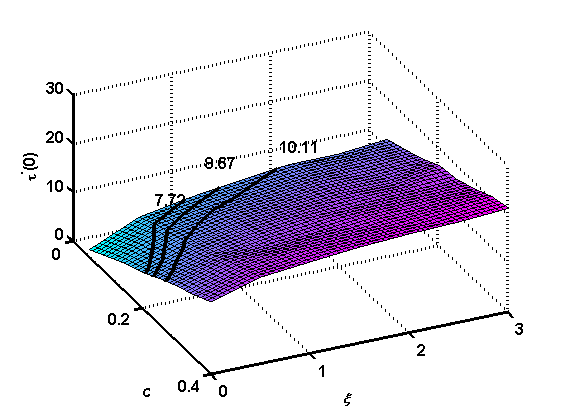}
\includegraphics[width=6cm]{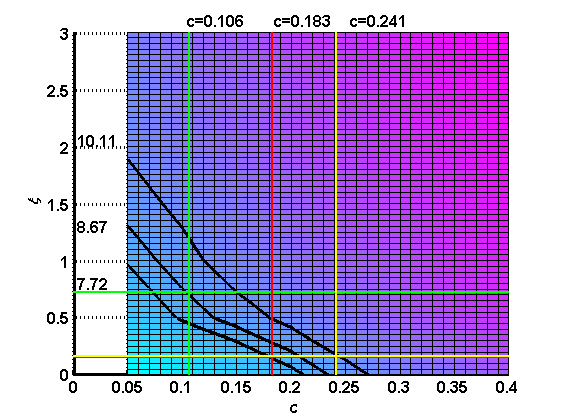}
\caption{(Colour online) Normalized ERF derivative at $\alpha =0$ plotted $versus$ NP concentration and SRO parameter for a collection of gold NPs on square lattice. Level curves for three experimental values of $c$ are shown. The parameters are the same as in Fig. \ref {tran:fig2}.} 
\label{tran:fig3}
\end{figure}

The calculated results are shown in Fig. \ref {tran:fig3}. First, in order to check the validity of the extrapolation procedure described above, it was applied to the calculated ERF spectra treated as "experimental" ones in this case. We verified that the SRO parameter obtained by applying this procedure coincides with the true one (used in the MC method), with a relative precision better than 1\%. We also applied the extrapolation procedure with regularization to real experimental spectra of the extinction coefficient \footnote {The extinction coefficient is related to the extinction cross-section by $\kappa =\sigma _e \lambda /(4\pi V)$, where $V$ is the volume of the system.} 
measured for thin colloidal gold films \cite {Langmuir2002}. Although their colloidal layers were not exactly 2D NP systems (they consisted of approximately two NP monolayers), this should not be a serious problem, because we use {\it normalized} spectra. 
The mapping of this system to our lattice model is as follows. (i) Assuming that NPs can touch, we take $l=2a_{NP}$, where $a_{NP}$ is the "physical NP radius" determined by means of transmission electron microscopy \cite {Langmuir2002}. Note that $a_{NP}>a$ since the NP's Au core was covered with a protecting shell. (ii) The Drude model parameters were either taken as for bulk gold ($\omega _p,\:\epsilon _\infty$) or evaluated using the experimental spectra of Ref. \cite {Langmuir2002}. Namely, from the broadening of the extinction spectrum of NPs in colloidal solution we obtained $\Gamma _p\approx 0.45$ eV and from the peak position in the spectrum of the most diluted film we found the "host dielectric constant", $\epsilon _h\approx 5.8$. Note that the latter is an effective parameter, since the particles located at the substrate/air interface and in each NP its metallic core is surrounded by its shell. Using a known semi-empirical relation \cite {Marzan2001} between $\Gamma _p$ and the metallic particle radius, we can estimate the core radius as $a \approx 3$ nm. Therefore, we used $a/l=0.25$ when computing the calibration data presented in Fig. \ref {tran:fig3}. 

The values of the "clustering" criterium for three NP films studied in Ref. \onlinecite {Langmuir2002} were calculated by roughly approximating the experimental spectra with Lorentzians and using the extrapolation procedure with regularization. They are shown in Fig. \ref {tran:fig3} defining "compatibility curves" $(c,\xi)$. If $c$ is known experimentally, the clustering parameter is determined. For the present example,     
the fraction of occupied sites can be related to the NP volume fraction $f$ (provided in Ref. \onlinecite {Langmuir2002}), by $c=6f/\pi$. This way we obtained the values of $c=0.11;\:0.18;\:0.24$ for the three films, respectively. Thus, we can conclude from Fig. \ref {tran:fig3} that there was very little clustering in these films, excepting perhaps the lowest NP concentration one ($\xi =0.73$), as qualitatively confirmed by their microscopy images \cite {Langmuir2002}.

In summary, we described a mathematical procedure allowing for the evaluation of the correlation parameter, which is a characteristic of short--range pairing of the polarizable particles, from the spectrum of an experimentally measurable quantity $\Im m (\alpha_0 \phi(\alpha_0))$ (normalized to its maximum value). To demonstrate the correctness of this procedure, we numerically generated "experimental" spectra of a model 2D system of Drude particles assuming the nearest-neighbour pairing with a predefined SRO parameter ($\xi $) and obtained its value with high precision using the extrapolation method. For a true experimental spectrum, the calculation of $\tau ^\prime(0)$ yields a set of compatible pairs of parameters, $(c,\:\xi)$, and, if the concentration is known independently, one determines $\xi$, as exemplified for the experimental data of Ref. \onlinecite {Langmuir2002}.
For an {\it isotropic} 3D ensemble of particles, the proposed criterium is shown to vanish, independently of the presence or absence of correlations. In this case, any significant deviation of $\tau ^\prime(0)$ from zero would imply an anisotropy in the system.
The knowledge of the single particle polarizability is required for the proposed analysis, in particular, the particle radius is important. Here we considered metallic NPs described by the Drude model but the procedure can be applied also to semiconductor quantum dots with the polarizability determined by localized exciton resonances.

{\it Acknowledgments.}
Financial support from the Portuguese Foundation for Science and Technology (FCT) through Projects PTDC-FIS-113199-2009. PEst-C/FIS/UI0607/2011 and PEst-C/MAT/UI0013/2011 is gratefully acknowledged.


\end{document}